\documentclass[epjCONF]{svjour}
\usepackage{graphics}
\usepackage[varg]{txfonts} % Times fonts
\usepackage[latin1]{inputenc}
\session-title{International Conference on New Frontiers in Physics 2012}
\begin{document}
\title{Some Aspects of String Cosmology and the LHC}
\author{Nick E. Mavromatos
%\fnmsep
\thanks{\email{Nikolaos.Mavromatos@kcl.ac.uk}} }
%\and Second author\inst{2} \and ... }
%
\institute{King's College London, Department of Physics, Strand, London WC2R 2LS, UK and  \\ CERN,  Physics Department, Theory Division, CH-1211 Geneva 23, Switzerland }
\abstract{
I discuss some (unconventional) aspects of String Cosmology of relevance to supersymmetric dark matter searches at the Large Hadron Collider (LHC) at CERN.
In particular, I analyse the r\^ole of time-dependent dilaton fields in relaxing some of the stringent constraints that characterise minimal supersymmetric models in standard cosmology. I also study briefly CPT-violating aspects of brane Universe models with space-time brane defects at early epochs and their potential relevance to the observed Baryon Asymmetry. 
} %end of abstract
\maketitle
\section{Introduction: Motivation and Summary}
\label{intro}
We are already into the second decade of the 21st century, and for the past twelve years we have been witnessing extraordinary developments and discoveries, both at the technological and conceptual levels,  in the fields of astro-particle physics and cosmology. 
The confirmation by a plethora of astrophysical experiments of the energy budget of our Universe, pointing to the fact that more than 70\% of it consists of a ``dark component'' of unknown nature (``dark energy''), and that more than 90\% of its matter component is also not understood at all (``dark martter''), have already changed our view of the cosmos from the one we have had in the previous century. On the other hand, at the particle physics front, the recent 
discovery at the Large Hadron Collider of CERN of a Higgs-like particle state~\cite{higgs}, probably puts together the final pieces in the puzzle of understanding the three out of four fundamental forces of Nature, electroweak and strong, and the origin of mass. All of this seems to  
be now well understood at low energies within the framework of the Standard Model (SM) of Particle Physics.
 The latter, however, fails to explain several aspects of our Universe: first, the observed (in the neutral Kaon system) CP violation does not suffice, if calculated within the SM framework and extrapolated at the early epochs of our Universe, to account for the observed matter-antimatter asymmetry or, equivalently, Baryon Asymmetry,  in the Universe (BAU)~\cite{cpbau}. One needs extra sources of CP violation, which can provide the necessary asymmetry in particle vs antiparticle decays in the early Universe, leading to the observed matter-antimatter asymmetry today, that is explaining why the Universe is mostly made of matter. This question is, therefore, of fundamental importance, as it relates to our very existence. 
Second, the SM matter is insufficient to account for the `missing mass' in the Universe, leading to its `dark matter' sector. Moreover, the SM \emph{per se}, being formulated without gravity,  does \emph{not} and it \emph{cannot}  provide an explanation for the Universe`dark vacuum energy' content. 
It is clear that in order to be able to answer these questions one should go Beyond the SM (BSM). Several research avenues, with diverse content and philosophy, exist at present, but none of them has given any definite answers nor have they been verified, as yet, by experiment. 
For instance, to tackle the issue of the understanding of the observed BAU, extensions of the SM which include supersymmetry~\cite{susy}, and/or sterile (right-handed) neutrinos~\cite{sterile}, which have no SM interactions and have suppressed mixing with the SM matter, have been proposed, since such theories do provide the required extra sources of CP Violation. As yet, there is no experimental evidence for both frameworks. 

For Supersymmetry (SUSY), minimal extensions of the SM, with partners of masses at TeV scale seem to have been on the verge of being ruled out by the current 
run at 8 TeV of LHC experiments (ATLAS and CMS)~\cite{susysearchlhc}. Nevertheless, there is currently a window that may survive the full run of LHC at 14 TeV and lead to a discovery. Notably, the recent LHC discovery of a Higgs-like state with mass around 126 GeV~\cite{higgs} can still accommodate such extensions~\cite{posthiggssusy}. In fact, it is the cosmology, that imposes severe restrictions in the available parameter space of the constrained minimal supersymmetric standard model (otherwise known as minimal SuperGravity (mSUGRA))~\cite{wmapsusy}~\footnote{Embedding SUSY in curved space-time, makes supersymmetry a gauge symmetry, and gravity is supersymmetrised, with the spin 3/2 fermion, the gravitino, playing the r\^ole of the partner of the graviton. 
Such embeddings restrict enormously the parameter space of SUSY models to just five parameters.}. 
This should be combined with the mass of the Higgs at 126 GeV  (which most-likely is the state that has been discovered at the LHC~\cite{higgs}), as well as  with  constraints 
coming from the muon gyromagnetic ratio ($g-2$) experiments vs theoretical models~\cite{wmapsusy}. 
The main restrictions from the cosmology come from the upper limits of the dark matter densities at the present era (normalised to the critical density of the Universe), upon the assumption that  \emph{minimal} supersymmetric models provide the leading candidate (single species) for (thermal) dark matter, in particular the neutralino. 

Going beyond minimal SUSY is  phenomenologically a highly complicated and model dependent process. Several supersymmetric models beyond mSUGRA do exist in the literature, which lead to entirely different phenomenology for dark matter than the minimal model and are still not ruled out by the current LHC run. R-partity violating models is one such class~\cite{rpv}.
Non-thermal dark matter species are also predicted in other models, which can avoid the stringent WMAP constraints. Example of such theories are heterotic string models on orbifolds, with unconventional breaking of low-energy supersymmetry~\cite{heterotic}. In general. string theory~\cite{strings} changes completely our view of the quantum physics, not only because it necessitates higher-dimensional background spaces, but mainly because the uniqueness of the vacuum which was a basic feature of  phenomenologically realistic quantum field theories, such as the Standard Model, seems to have been lost. There is a \emph{Landscape } of (mathematically) allowed string theory vacua, and at present it seems that we do not have a principle of distinguishing them and thus pick up one preferentially. Minimisation of energy does \emph{not} work for this purpose. 
This lead~\cite{landscape} to ``anthropic principle-like'' arguments for the ``selection'' of the physical string vacuum that incorporates the SM as we know it, which, however, personally I do not find satisfactory. Hence it remains to be seen whether string theory is a true physical theory of Nature. So far it resists any concrete experimental predictions and of course experimental evidence, but these are not necessarily drawbacks, because they are not unique to strings. As we have seen, both supersymmetry and sterile neutrino models suffer from similar problems. Nevertheless, string theory is an intellectually challenging and mathematically consistent framework for discussing fundamental issues of space-time, such as the origin of the Universe and quantisation of gravitational interactions in unison with  the rest of the fundamental interactions in Nature. In this respect, it worths attempting to answer the above-described questions, concerning the nature of the dark sector of the Universe and the observed Baryon-Asymmetry, within this framework. 
However, this may be too difficult a task. Our knowledge or understanding of strings (and their extensions, brane theory) are at present quite limited, despite the fact that more than twenty years of intense theoretical work have been devoted to it and spectacular progress has been achieved.
Nonetheless, attempts to pick up certain features, unique to strings, that might characterise low-energy string-inspired extensions of the SM, may indeed lead to breakthroughs in our understanding without the need to fully comprehend  quantum string theory at microscopic scales. In fact, picking up the minimal extensions of SM that capture such more-or-less unique to strings features, may provide ways of tackling, for instance, the above-mentioned problems regarding supersymmetric searches at colliders and avoiding stringent cosmological restrictions, as a result of string theory changing the standard cosmological scenarios in a minimal but crucial way. 
It is the point of this talk to discuss how such minimal extensions of the SM, inspired from string theory, do provide a way out of these problems, and also how they can account for the observed baryon asymmetry in the Universe.

The structure of the talk is as follows: in the next section \ref{sec:1} we concentrate on such a feature of string-inspired cosmological models, involving time-dependent cosmic dilaton fields, and argue how the thermal relic abundance of dark matter candidates of mSUGRA-like models, when embedded in this framework, is diluted, compared to standard cosmology with no (or constant) dilatons, to a point where the current LHC/WMAP constraints can be avoided. 
In the following section, sec.~\ref{sec:2}, I discuss first how CPT symmetry may break in the lepton (neutrino) sector of such string-inspired models in the Early Universe, due to different dispersion relations between leptons and antileptons, induced by their propagation in a Lorentz-Violating `medium' of space-time brane defects. This then leads to lepton-antilepton asymmetries, that freeze out at the high temperatures of the neutrino decoupling, and are then communicated to the baryon sector through Baryon-Lepton-number (B-L) conserving processes in grand unified string-imspired models, thus providing an explanation of the 
observed BAU. Conclusions and Outlook are presented in section \ref{sec:3}. It must be stressed, that the contents of this article reflect personal research interests of the speaker and therefore they by no means constitute a comprehensive review of the subject of string cosmology. 

\vspace{-0.4cm}

\section{Dilaton-driven String-Inspired Cosmology and the LHC}
\label{sec:1}

String-inspired cosmology is a  local field theory, which is obtained as a low-energy limit of some string theory model of the Universe.
It is \emph{assumed} here that string theory has settled in one of its ground states, or at least approaches it, and that this ground state is our observable Universe. 
Thus string-inspired Cosmology does \emph{not} attempt to give a solution to the Landscape problem, mentioned earlier, but rather bypasses it altogether and concentrates on specific predictions, especially as far as the dark sector of the Universe is concerned. The latter can in principle be tested by means of the astrophysical experiments available to date or in the foreseeable future. 

\subsection{String Moduli Fields}\label{subsec:1} 

One of the distinguishing characteristic features of string models is the plethora of the so-called \emph{moduli} fields, that is elementary scalar excitations, 
with continuous families of global minima.Their expectation values \emph{label} various string backgrounds. 
String moduli of interest to us in this work, include: 
\begin{itemize} 
\item{} Scalar fields associated with extra space dimensions, with \emph{no} classical  potentials, \emph{i.e.} the latter may be generated by quantum string-loops, and are therefore suppressed by (positive) powers of the string coupling $g_s$. These moduli fields have Planck-scale suppressed couplings to SM matter.
Their expectation values classically describe the size and configuration of the extra space dimensions that characterise string and/or brane models. Moduli stabilisation is an open issue in many string models. Models with stabilised moduli, which are massive, do exist~\cite{acharya}. The moduli masses can be of order of the string mass scale $M_s$. In string theory, at least as we understand it today, $M_s$ is a free parameter, which can be in the range ${\rm O}(10)~{\rm TeV} \le M_s \le 10^{15} ~{\rm TeV} = M_P/10$, with $M_P$ the four-dimensional Planck mass. The lower bound is phenomenologically-imposed (by the current run of LHC experiments at 8 TeV centre-of-mass energy ).  In concrete string models~\cite{acharya}, 
such massive moduli can decay before the Big-Bang-Nucleosynthesis (BBN) and can thus act as significant ``sources'' of \emph{non-thermal} Dark Matter (DM), having important effects on collider searches for supersymmetry, by avoiding most of the stringent constraints of models with thermal DM relics in conventional Cosmology. 

\item{} The Dilaton Field, $\phi$, whose \emph{stabilised} vacuum expectation value $<\phi >$ is associated with the 
value of the string coupling, 
\begin{equation}\label{stringcoupl}
g_s = {\rm exp}\big(\phi\big)
\end{equation}
which in turn determines, upon compactification, the four-dimensional gauge couplings and thus the low-energy phenomenology of the associated string-inspired particle-physics field theory models. In this talk, we shall consider \emph{non-equilibrium} Stringy Cosmologies, characterised by the \emph{dominance } of 
non-stabilised, \emph{i.e}. cosmic time dependent (``r\emph{unning}'') Dilatons~\cite{gasperini} at early epochs before the Big Bang Nucleosynthesis (BBN). 
Such running dilaton dominance may or may not, depending on the model, last until the present era. As we shall discuss below, the effect of such running dilatons is to \emph{dilute}~\cite{lahanas} the available thermal dark matter relics that couple to it. In this way, the stringent constraints, imposed by cosmological considerations on the constrained minimal supersymmetric models, for instance, in the framework of standard cosmologies,  are avoided. Much more room for minimal supersymmetry is allowed by cosmology in the relevant parameter space of such models~\cite{dutta,spanos}, in agreement with the current LHC data. 
It should be mentioned here that Dilaton Cosmologies have been considered in the literature so far from a different point of view, in particular providing ways to study non-perturbatively pre-Big-Bang scenarios in the history of our Universe~\cite{pBB,gasperini}, with interesting early Universe phenomenological signatures, especially in the spectrum of primordial fluctuations. Although very interesting and thought stimulating, unfortunately such aspects 
will not be discussed in the talk, because of lack of space and time. As already mentioned, we shall exclusively concentrate here on the effects of running dilaton on the thermal DM abundances in minimal supersymmetric models and the associated collider searches, which we now proceed to analyse.

\end{itemize}

\subsection{Running Dilatons and Dilution of Thermal DM relics }\label{subset;2}

Before understanding qualitatively the effects of running dilaton fields on thermal DM relics, it would be instructive to motivate their existence within the framework of string theory. As already mentioned in the previous subsection, most of the phenomenology of strings is based on the fact that the four-dimensional dilaton fields, which are the scalar (spin zero) part of the \emph{massless} gravitational excitations of the string, are stabilised in their expectation values, somehow.
Such stabilised dilatons are therefore space-time independent. The mechanism for such a stabilisation is still unknown, given that the dilaton potential, which vanishes at tree level, receives contributions from higher string loops, that are not well-understood at present and probably include stringy non-perturbative effects. 

\begin{figure}[tb]
\begin{center}
\resizebox{0.35\columnwidth}{!}{%
  \includegraphics{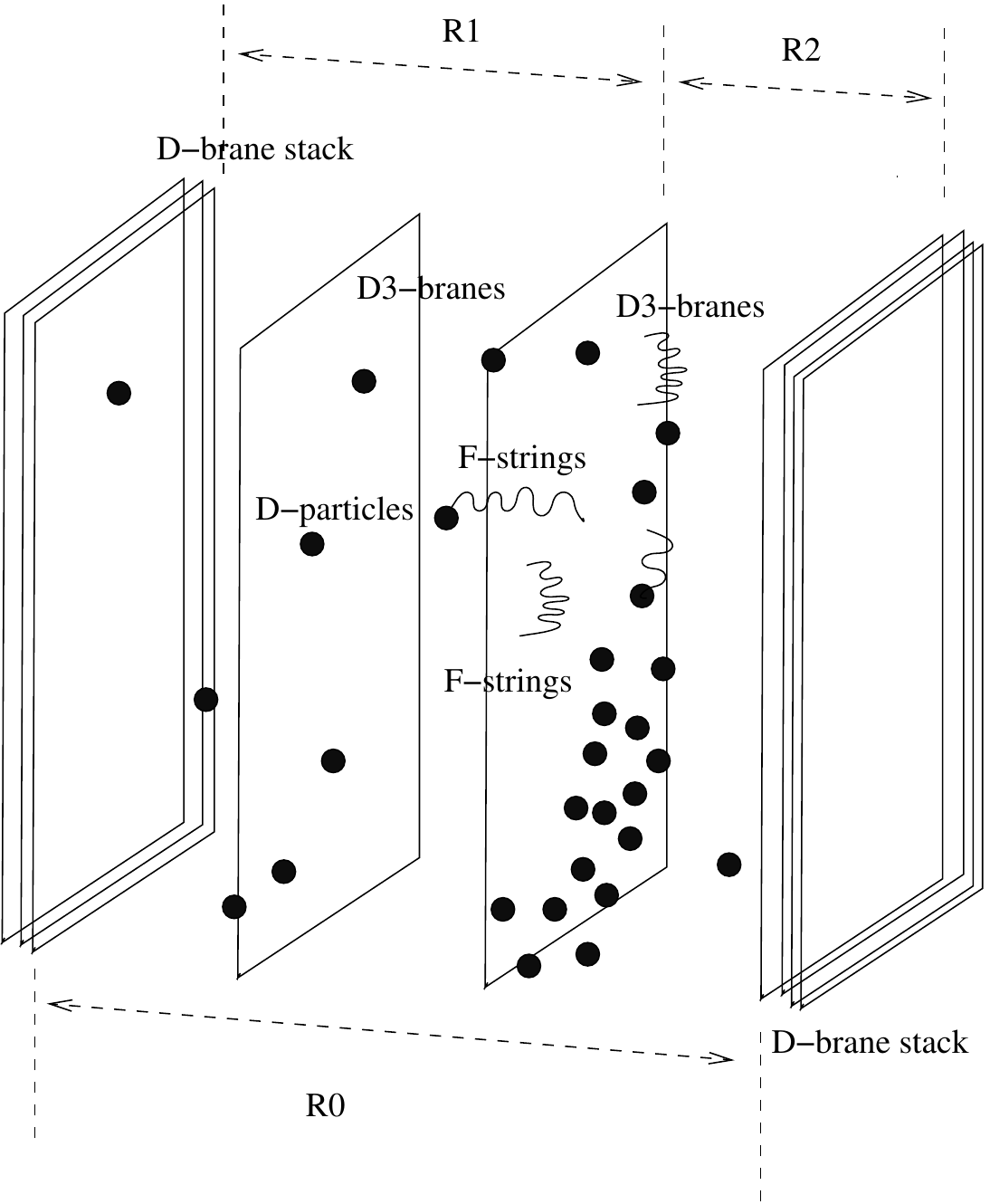} }
  \end{center} 
\caption{A Bouncing brane scenario can provide a concrete example on how time-dependent dilaton fields may enter certain non-equilibrium phases in the early Universe history. During the bounce of two of the brane worlds (labelled D3-branes), the extra-dimensional bulk separation depends on the cosmic time, and the latter implies a non trivial time-dependence on the Dilaton fields. The black dots in the figure denote possible D-particle defects, that is compactified D-branes of lesser dimensionality than the brane worlds, which can provide extra sources of Dark-like matter (D-matter), but also CPT violation in the early Universe.}
\label{fig:brane}       % Give a unique label
\end{figure}

However, in the early universe, there may be (non-equilibrium)  phases, for instance before the BBN, where non constant, time dependent dilaton fields may dominate. A straightforward example of such a scenario is the aforementioned one on pre Big-Bang Cosmology, where at very early Universe epochs 
before and immediately after the Big bang (a cosmically catastrophic event that is not specified in these scenarios in detail) there is a non trivial dilaton potential.
The dilaton is then time dependent and under certain circumstances can climb up the pre-Big-Bang potential, which is characterised by a very small height at the time of the Big Bang,  until it passes a local maximum, after which time it starts rolling down towards a global minimum of the potential, where the dilaton field is stabilised.  In another, brane Universe inspired scenario, our Universe is one of two colliding branes that have been separated from two corresponding stacks of 
multiple coincident D(irichlet) branes~\cite{polchinski}, used to provide bulk-space  boundaries, and thus dynamical compactification~\cite{westmuckett}
(see fig. \ref{fig:brane}). The mechanism of the separation may be due, for instance, to quantum brane fluctuations. Each of these branes represents a Universe, and one of them our observable (D3-brane) Universe (after appropriate compactification to three spatial dimensions). 
Open strings, with their ends attached to the observable-Universe D3 brane, represent SM excitations on our world. Excitations in the Gravitational String Multiplet in this picture, such as gravitons and gravitinos, are allowed to propagate in the bulk and are represented by closed strings. 
The Big bang in this picture would correspond to a cosmically catastrophic event when these two D3-branes collide and  
bounce back.  For the purposes of this section the reader is invited to ignore the black blobs appearing in the bulk space between branes in fig.~\ref{fig:brane}. These denote compactified D-branes of smaller dimensionality than the number of longitudinal (uncompactified) dimensions of the brane worlds, which then look, from the viewpoint of  brane observers,   as \emph{effectively  point-like} objects/defects in the brane space-time (``\emph{D-particles'}'). These will play a role analogous to dark matter and may provide the seeds for CPT violation in the early universe, a topic we shall discuss in the next section.  The bulk distribution of D-particles need not be uniform. 

There are complicated potentials among the various D-branes in this scenario that eventually will stabilise the branes at certain positions in the bulk extra-dimensional space. However,  for a certain period of time after the bounce, the D3-branes will move away from each other, and during this non-equilibrium phase, which can be a phase before the BBN era on the observable Universe brane, there is a non trivial dilaton, depending on cosmic time. 
To understand this latter feature, the reader should recall the results of \cite{rizos}, pointing out to the existence of exact solutions of low-energy field theories with higher curvature corrections  in equilibrium brane scenarios with branes separated at  \emph{fixed}  bulk distance $z$, assuming for simplicity and concreteness just one large bulk extra dimension. The corresponding configurations for the dilaton field in such a situation, which are solutions of the five-dimensional gravitational and scalar equations of motion,  depend on the branes separation $z$ along the  bulk extra dimension, and are given by:
\begin{equation}\label{dilatonbulk}
\phi (z) = \phi_0 \, {\rm ln}|z| 
\end{equation}
with $\phi_0 < 0$ in the models of \cite{rizos}. For moving branes, the relative separation $z$ will be a function of the cosmic time. For instance, for adiabatically moving branes, with relative bulk velocity $U \ll c$, in a direction transverse to the longitudinal brane dimensions (\emph{cf.} fig. \ref{fig:brane}), the separation $z=U t + z_0 $, which in view of (\ref{dilatonbulk}) implies a situation where there are non-trivial running dilaton fields
\begin{equation}\label{dilatonbulktime}
\phi (z(t)) = \phi_0 \, {\rm ln}|U\, t + z_0|~. 
\end{equation}
This phase of course lasts until the branes are stabilised, when the dilatons settle in the minimum of their (unknown) potentials. This phase may be a brief phase in the history of the Universe, before the BBN, which however could affect DM relics by diluting their abundances in a way we now proceed to explain. 
Our considerations that follow will be rather generic, without making reference to specific models with time dependent dilatons. This suffices for giving the reader a rather generic and model independent description of the situation, which is fairly general for this sort of non-equilibrium cosmologies. 

The physics of running dilaton cosmology at low-energies, where only terms quadratic in space-time derivatives, including the Einstein  term, are taken into account  in the effective gravitational Lagrangian, is captured by the following cosmological equations of motion that replace the ones of the Friedmann-Roberetson-Walker Cosmology~\cite{gasperini}:
\begin{eqnarray}\label{dilatoncosmo}
{\ddot \phi} + 3 H \, {\dot \phi} + V^\prime (\phi) = 0, \quad 
3 H^2 = \frac{1}{2} \big({\dot \phi}\big)^2 + V(\phi) ~, \quad 
2{\dot H} = - (\rho_\phi + p_\phi) = - \big({\dot \phi}\big)^2~, 
\end{eqnarray}
and the continuity equation
\begin{equation}\label{contin}
\frac{d}{dt} \, \rho + 3 H (\rho + p) - \frac{{\dot \phi}}{\sqrt{2}} \big(\rho - 3 p \big) = 0~, 
\end{equation}
where in the above formulae, the overdot denotes time derivative,  $H$ is the Hubble parameter, $\phi$ is the running dilaton, depending on cosmic time only,  so that there is no problem with the observed large-scale isotropy and homogeneity of the Universe,  the $\rho, p$ are respectively the energy density and pressure of matter or radiation, and $\rho_\phi, p_\phi$ are the corresponding quantities for the dilaton fluid. 
The quantity $V(\phi)$ denotes the potential of the dilaton, which in string theory arises from string loops and, as mentioned previously, is not known in general.
Fortunately, its details will not matter for what we want to demonstrate below. Finally, a prime over $V$ denotes the functional differentiation of $V$ with respect to the dilaton field $\phi$.

We focus our attention on Eq.~(\ref{contin}), and in particular its ${\dot \phi}$-dependent term.  For any non-zero ${\dot \phi} \ne 0$ there will be extra contributions to the standard continuity equations for matter but not for radiation, since in the latter case this term vanishes identically due to the radiation equation of state 
$p=\rho/3$. Thus for (dark) matter, where $p=0$, one would obtain a modified Boltzmann equation for the evolution of its thermal relics. For instance, for the simplified case of a dominant DM species of mass $m$, the energy density $\rho = m \, n$, where $n$ denotes the number density of the species. 
The latter, would then obey the following Boltzmann evolution equation, following from (\ref{contin})~\cite{lahanas}: 
\begin{equation}\label{boltz}
\frac{d}{d \, t} \, n + 3 \, H \, n = \frac{{\dot \phi}}{\sqrt{2}} \, n  + \int \, \frac{d^3p}{E} \, C[f(\vec p, t)]~,
\end{equation}
where $n = \int d^3 p \, f( \vec p, t)$, with $f(\vec p, t)$ 
a phase-space distribution for DM species, $E$ is the energy of the species and  $C[f (\vec p, t)] $ is the Boltzmann Collision term.  
Depending on the sign of ${\dot \phi}$ one would have dilution or increase of the thermal DM relic abundance.

For instance, in the bouncing brane model, where the bulk separation between the branes \emph{increases} with the cosmic time for a give short period in the early Universe epoch, we have from (\ref{dilatonbulktime}) that ${\dot \phi} < 0 $, and so in this case there would be dilution of the DM thermal relic abundance. 
In detailed models coming from (non-critical) strng/brane theory, which could describe early Universe non-equilibrium Cosmologies~\cite{lahanasdetails}, it can be arranged that only DM abundances are affected by the diluting dilaton effects, while baryonic matter remains unscathed, in a way consistent with recent cosmological and astrophysical data (including galactic-growth data, in cases where the non-constant dilaton is present after BBN, during structure formation era~\cite{astro}). 

\subsection{Running-Dilaton Effects on Collider Searches for Minimal Supersymmetry}\label{susec:3}

\begin{figure}[tb]
\begin{center}
\resizebox{0.90\columnwidth}{!}{%
  \includegraphics{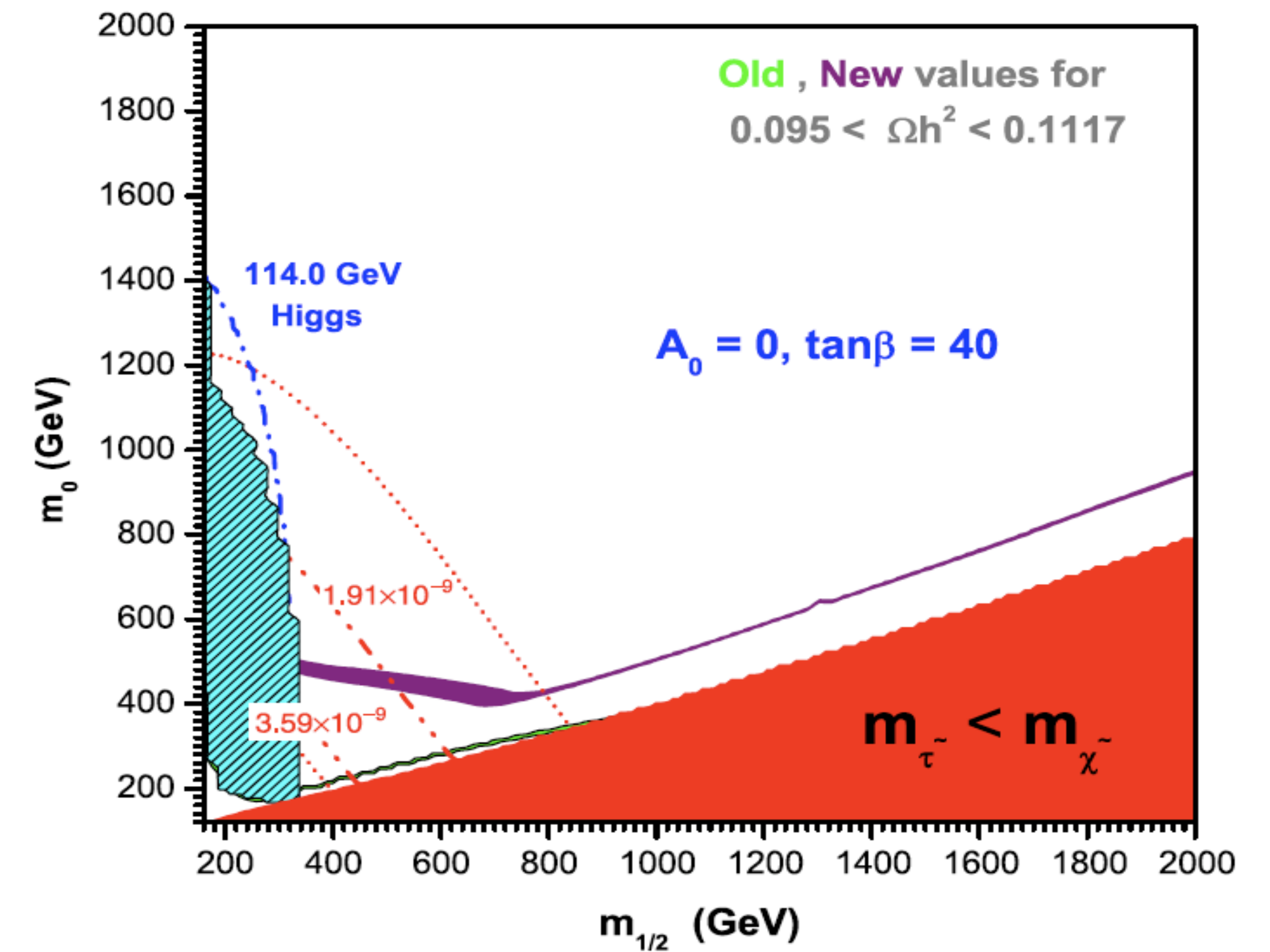}  \hfill     \includegraphics{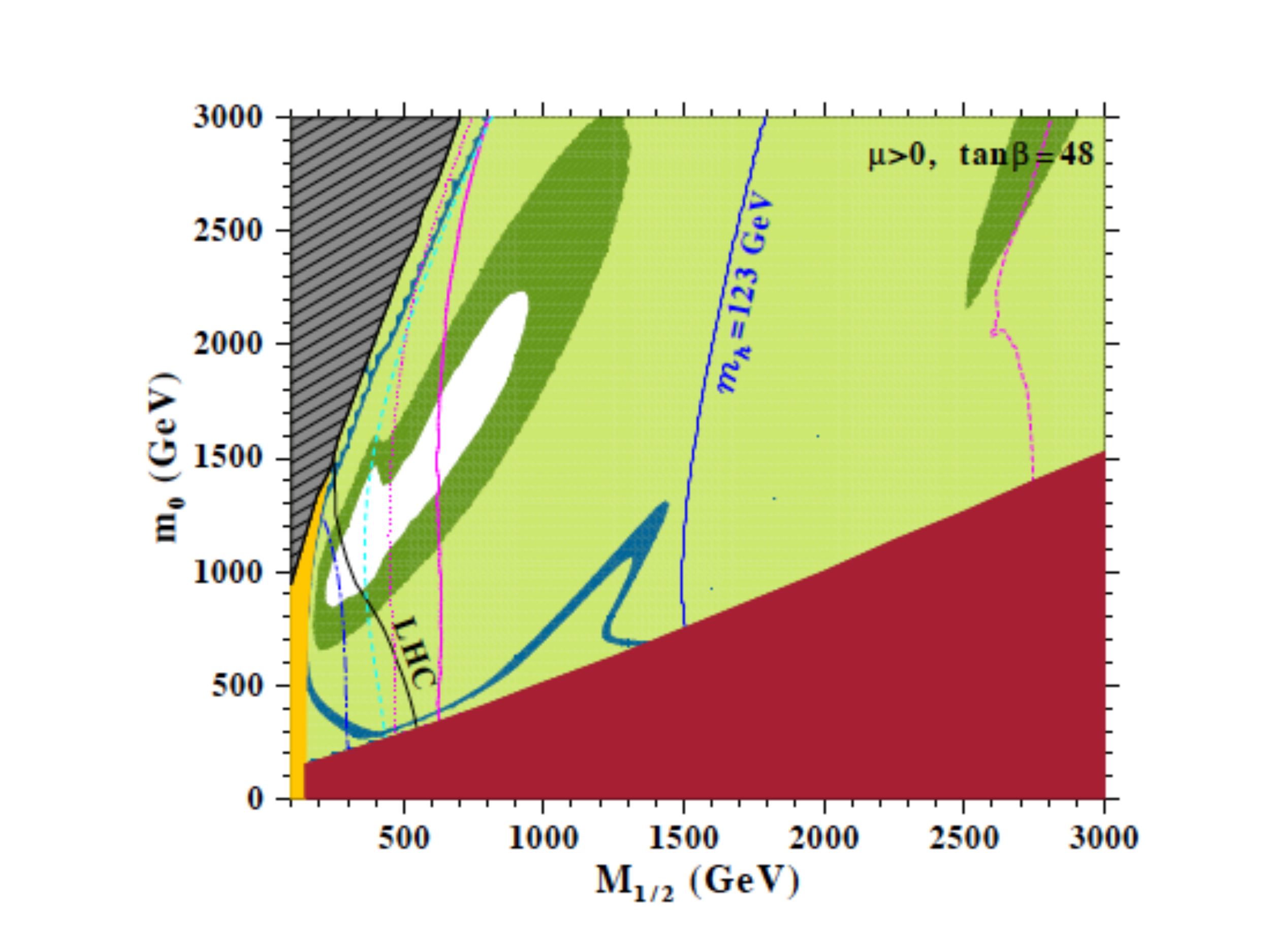}}
  \end{center} 
\caption{\emph{Left Picture}: Examples of an enlargement of the Cosmology-allowed parameter space for Constrained Minimal Supersymmetric Standard Model (mSUGRA), as a result of the dilution of the neutralino thermal DM abundances due to Running Dilaton effects (Old(standard, running-dilaton-free) Cosmology-allowed region, given by light-green shaded  strip. Running-Dilaton-Cosmology-allowed region is given by purpled coloured strips (from Ref.~\cite{lahanas})).
\emph{Right Picture: }
The enlargement can be significant, pushing Cosmology-allowed SUSY partner masses event beyond the reach of LHC experiments (Running-dilaton Cosmology-allowed region: dark-green shaded). From Ref.~\cite{spanos}, where a post-Higgs-like-LHC-discovery analysis has been made.}
\label{fig:susy}       % Give a unique label
\end{figure}

As discussed in \cite{dutta}, a dilution of order O(1/10) of DM relic suffices to create interesting phenomenology at collider searches of supersymmetry, in the sense that much heavier SUSY partners, than in conventional scenarios, are allowed now, due to the significant enlargement of the minimal SUSY parameter space (\emph{cf.}  fig.~\ref{fig:susy}). The presence of much heavier SUSY partners will affect the corresponding collider signatures of SUSY, say at the LHC, if this scenario is realised. 
For instance, the final states expected at the LHC in this case would consist of Z-bosons, Higgs Bosons and high-energy $\tau$-leptons, unlike the standard scenarios. Such states are produced when one looks at the decay chains of the dominant SUSY production mechanism of squark ($\tilde q$) and gluino ($\tilde g$) pairs at the LHC: 
$ \tilde q \, \rightarrow \, q \, \chi_2^0 \, \rightarrow \, q \tau \, \tilde \tau_1 \, \rightarrow \, q \tau \tau {\tilde \chi}^0_1 ~, \quad 
{\tilde \chi}^0_2 \, \rightarrow \, h^0\, {\tilde \chi}^0_1 ~, \quad {\tilde \chi}^0_2 \, \rightarrow \, Z\, {\tilde \chi}^0_1 ~, $
where  ${\tilde \chi}^0_1$ is the neutralino, assumed to be the dominant DM candidate in the running-dilaton-enhanced mSUGRA model of \cite{dutta}, 
${\tilde \chi}^0_2$ is the next-to-lightest neutralino, and $h^0$ is the Higgs particle. A detailed analysis in the standard parameter space $m_{1/2}- m_0$ (with $m_{1/2} (m_0)$ the gaugino (saclar) mass parameter) of mSUGRA 
models has been performed in \cite{dutta}, indicating the regions where each of the above decay patterns is dominant and the corresponding signals at the 14 TeV run of LHC, which can be summarised in the following: 
\begin{itemize} 
\item{(i)} \, Higgs + jets + missing~transverse~energy, 
\item{(ii)}  \, Z + jets + missing~transverse~energy, 
\item{(iii)}\, 2$\tau$ +  jets + missing~transverse~energy~. \nonumber 
\end{itemize}

Moreover, as discussed in \cite{spanos}, in an updated post-Higgs-like-LHC discovery analysis of such running dilaton scenarios, for larger dilaton-induced dilutions of the DM,  a light Higgs of 126 GeV mass can be made compatible with such dilaton-affected minimal SUSY models, in the sense that there are cases where the dilaton-cosmology allowed regions of SUSY can even be pushed beyond the reach of LHC for large tan$\beta > 48$. 

Hence, we observe that the simplest extension of the minimal supersymmetric model, obtained by embedding it to a string-inspired cosmological framework with a running dilaton at early epochs, suffices to enlarge the Cosmology-allowed parameter space, and evades the stringent constraints imposed on minimal SUSY in conventional dilaton-free Cosmologies. Of course, to construct detailed cosmologies with such running dilators is an open and model-dependent issue. The lack of knowledge of a dilaton (and in general other string moduli) potential hampers a complete treatment of the problem, restricting any potential analyses at present to phenomenological ones, such as the one above.  Nevertheless it is hoped that the progress that has been made in recent years towards an understanding of non-perturbative stringy effects, especially in the context of brane Universe models, can improve the situation in the foreseeable future. 

\section{CPT Violation in the Early Universe in String/Brane-Inspired Cosmology and Baryon Asymmetry}
\label{sec:2}

We now move to discuss the potential r\^ole of string theory in answering another important open issue of modern astro-particle physics and Cosmology, namely the origin of the observed dominance of matter over antimatter in the Universe. 
We remind the reader that the almost 100\% observed asymmetry today, is estimated in the Big-Bang theory~\cite{Gamow} to be of order: 
\begin{equation}\label{basym}
\Delta n (T \sim 1~{\rm GeV}) = \frac{n_{B} - n_{\overline{B}}}{n_{B} + n_{\overline{B}}} \sim \frac{n_{B} - n_{\overline{B}}}{s} = (8.4 - 8.9) \times 10^{-11} 
\end{equation}
at the early stages of the expansion, e.g. for times $t < 10^{-6}$~s and 
temperatures $T > 1 $~GeV. In the above formula $n_B$ ($n_{\overline{B}}$) denotes the (anti) baryon density in the Universe, and $s$ is the entropy density.  

As already mentioned in the introduction, the standard sources of CP violation within the framework of the SM of particle physics are not sufficient to reproduce the above asymmetry~\cite{cpbau} and hence one should look for alternative sources beyond the SM. 
String theory, being inherently a supersymmetric theory of space-time, includes in its low energy limit extra sources of CP violation and in principle it can reproduce this number. The literature is vast and will not be covered here. 
In the current talk I would like to focus on the specific bouncing brane theory model considered previously (\emph{cf.} fig.~\ref{fig:brane}) and discuss the r\^ole of D-particle defects on inducing a lepton/anti-lepton asymmetry in the early universe, and through B-L conserving processes the observed BAU, without making specific assumptions about the supersymmetry content of the model. If true, this mechanism then of defect-induced BAU would be a completely novel feature of string/brane theory, which - in view of the discussion in the previous section - could also account for a potential absence of any  SUSY signals at the LHC energy scales in a natural and self-consistent way.

In this scenario there is a preferential r\^ole of neutrinos in feeling the full effects of the 
D-particle ``medium'' (\emph{cf}. fig.~\ref{fig:brane}), and hence the CPT Violation (CPTV), as we shall discuss below, which is attributed
to electric charge conservation: the representation of SM particles
as open strings, with their ends attached to the brane worlds, prevents
capture and splitting of open strings carrying electric fluxes by
the D-particles (the reader should recall that in string theory the electric
charge is at the end point of an open string). D-particles are electrically
neutral and thus electric charge would not have been conserved if
such processes had taken place. Hence,
the D-particle ``medium''  is transparent to charged excitations of the SM,
leaving neutral particles, in particular neutrinos, to be vulnerable
to the D-particle-medium effects. As discussed in detail in \cite{Mavromatos:2010nk} the density of
D-particles on the brane world can be relatively large, even at late
eras of the Universe, given the fact that bulk D-particles exert attractive
forces on the brane Universe with mixed sign contributions to the
brane vacuum energy, depending on the distance of the bulk D-particles
from the brane. Such forces are due to stretched strings between the
defect and the brane. These energy contributions depend only on the
transverse to the brane worlds components of the relative velocities
of the defect with respect to the brane. Thus, issues such as overclosure
of the Universe by a significant population of D-particle defects
can be avoided. For our purposes in this work we may therefore consider
that statistically significant populations of D-particles existed
in the early eras of the brane Universe. As the time elapses, the
brane Universe, which propagates in the higher-dimensional bulk (\emph{cf.}
fig.~\ref{fig:brane}), enters regions characterised by D-particle
depletion, in such a way that the late-era cosmology of the Universe
is not affected. Nevertheless, as we shall discuss below, the early
D-particle populations may still have important effects in generating
neutrino-antineutrino populations differences (asymmetries), which
are then communicated to the baryon sector via the standard sphaleron
processes~\cite{cpbau} or B-L conserving GUT symmetries in unified particle physics
models. To this end, we need to consider the effective dispersion relation
of a (anti)neutrino field in a brane space-time punctured with statistically
significant populations of D-particles. The latter is a dynamical
population, consisting of defects crossing the brane all the time,
thereby appearing to a brane observer as flashing ``on'' and ``off''
space-time ``foamy'' structures. The (anti)neutrino excitations
are represented as matter open strings with their ends attached on
the brane.The number density of (anti) neutrinos on the brane world
is limited by the requirement that they do not overclose the Universe.
Standard cosmological neutrino models predict the number densities
of a single flavour of relativistic neutrinos \emph{plus}
antineutrinos in thermal equilibrium at temperature $T_{\nu}$.
In order for our calculations below to make sense we should then assume that at the early Universe 
era of neutrino decoupling, when we are interested in calculating the induced CPTV, the density of D-particles is much higher than the corresponding one of (anti) neutrinos.   This can be arranged in specific models~\cite{sarbennick}. 

We now proceed to estimate the modification
of the dispersion relations of neutrinos in such ``media'' of D-particles
in the early universe. Interaction with the D-particles implies that
at least one of the ends of the open string representing the neutrino field is attached to the D-particle defect, with
the simultaneous creation of virtual strings stretched between the
defect and the brane, that describe the recoil of the D-particle.
During the interaction time, the D-particle undergoes motion characterized
by non trivial velocities, $u_{\parallel}=\frac{g_{s}}{M_{s}}\Delta k_{i}=\frac{g_{s}}{M_{s}}r_{i}\, k_{i}$
along the brane longitudinal dimensions, where $r_{i}$ denotes the
proportion of the incident neutrino momentum that corresponds to the
momentum transfer $\Delta k_{i}$ during the scattering. 
We have also assumed that the fraction of the neutrino momentum transfer
in the direction perpendicular to the brane world is negligible.  Meanwhile, other bulk D-particles (\emph{cf.} fig.~\ref{fig:brane}) exert forces on the vacuum energy
of the brane world of mixed sign, depending on their relative distance.
Thus, during the scattering process of a neutrino field with a D-particle,
the vacuum energy of the brane fluctuates by an amount $\Delta\mathcal{V}$
which depending on the process can be of either sign. From energy-momentum
conservation, at each individual scattering event between a neutrino
field and a recoiling D-particle, one could thus write:
\begin{equation}
\vec{p_{{\rm before}}}+\vec{p}_{{\rm after}}+\frac{M_{s}}{g_{s}}\,\vec{u}_{\parallel}=0~,\quad E_{{\rm before}}=E_{{\rm after}}+\frac{1}{2}\,\frac{M_{s}}{g_{s}}\,{\vec{u}}_{\parallel}^{2}+\Delta\mathcal{V}\label{neutrino}
\end{equation}
 where $(\vec{p},\, E)_{{\rm before\,(after)}}$ denote the incident
(outgoing) neutrino momenta, energies repectively and we used the
fact that the recoiling heavy D-particle of mass $M_{s}/g_{s}$ (with
$M_{s}$ the string scale and $g_{s}<1$ the string coupling, assumed
weak, so that string perturbation theory applies) has a non-relativistic
kinetic energy $\frac{1}{2}\,\frac{M_{s}}{g_{s}}\,{\vec{u}}_{\parallel}^{2}$.
Upon averaging $\langle\langle\dots\rangle\rangle$ over a
statistically significant number of events, due to multiple scatterings
in a D-particle-foam background, we may use the following stochastic hypothesis~\cite{Bernabeu:2006av}
\begin{equation}
\ll u_{i\,\parallel}\gg=0~,\qquad\ll u_{i\,\parallel}u_{j\,\parallel}\gg=\sigma^{2}\delta_{ij}~, \qquad \ll \Delta\mathcal{V}\gg = 0~, \label{LI}
\end{equation} 
implying that Lorentz invariance holds \emph{only} as an \emph{average} symmetry
over large populations of D-particles in the foam. 
It is this violation of LI in \emph{stochastic fluctuations} (or, equivalently, \emph{ locally} at individual scatterings of stringy matter off D-particles, due to the recoil of the latter) that is associated with the induced CPT Violation, in the sense of 
different dispersion relations between particles and antiparticles induced by the recoiling D-particle medium. 
On averaging over populations of D-particles, we need to find an expression for $\ll E_{\rm before}\gg $ appearing in (\ref{neutrino}). 
To this end, we mention that the effects of the D-foam go beyond the above-mentioned kinematical
ones. As discussed in ~\cite{Ellis:1999uh,Bernabeu:2006av} the non-trivial
capture and splitting of the open string during its interaction with
the D-particle, and the recoil of the latter, result in a \emph{local}
effective space-time metric distortion, in the neighbourhood of the recoiling D-particle, of the form:
\begin{equation}
ds^{2}=g_{\mu\nu}dx^{\mu}dx^{\nu}=(\eta_{\mu\nu}+h_{\mu\nu})dx^{\mu}dx^{\nu}~,\qquad h_{0i}=(u_{i\,\parallel}^{a}\sigma_{a})~,\label{recmetric}
\end{equation}
 where $u_{i\,\parallel}$ is the recoil velocity of the D-particle
\textit{on} the D-brane world, with $i=1,2,3$ a spatial space-time
index, $\sigma_{a}$ are the $2\times2$ Pauli flavour matrices with
$a=1,2,3$ (assuming two-neutrino-flavour oscillations for simplicity). On
average over a population of stochastically fluctuating D-particles
including neutrino-flavour changes, one may have the conditions (\ref{LI}),
the second of which, in the case of flavour oscillations, can be generalised
to $\ll u_{a,i}^{\parallel}u_{b,j}^{\parallel}\gg=\sigma^{2}\delta_{ij}\delta_{ab}.$
(we still assume that $\ll u_{a,i}^{\parallel}\gg=0$ . ) As a result
of this, on average, the flavour change during the interactions
of neutrinos with the D-foam can be ignored. In such a case, any flavour
structure in the metric (\ref{recmetric}) is ignored. On assuming
isotropic momentum transfer, $r_{i}=r$ for all $i=1,2,3$. The dispersion
relation of a neutrino of mass $m$ propagating on such a deformed
isotropic space-time, then, reads:
$p^{\mu}p^{\nu}g_{\mu\nu}=p^{\mu}p^{\nu}(\eta_{\mu\nu}+h_{\mu\nu})=-m^{2}\,\Rightarrow\, E^{2}-2E{\vec{p}}\cdot u_{\parallel}-{\vec{p}}^{2}-m^{2}=0.$
This on-shell condition implies that
$E=\vec{u_{\parallel}}\cdot{\vec{p}}\pm\sqrt{({\vec{u}}\cdot{\vec{p}})^{2}+{\vec{p}}^{2}+m^{2}}~.$
We take the average $\ll\dots\gg$ over D-particle populations with
the stochastic processes (\ref{LI}). Hence we arrive
at the following expression for an average neutrino energy in the
D-foam background:
\begin{eqnarray}
\ll E\gg  =  \ll\vec{p}\cdot{\vec{u}}\gg\pm\ll\sqrt{p^{2}+m^{2}+({\vec{p}}\cdot{\vec{u}})^{2}}\gg
  \simeq  \pm\sqrt{p^{2}+m^{2}}\left(1+\frac{1}{2}\sigma^{2}\right),\qquad p\gg m~,\label{dispersion3}
\end{eqnarray}
for the active light neutrino species. The last relation in eq.~(\ref{dispersion3})
expresses the corrections due to the space-time distortion of the
stochastic foam to the free neutrino propagation. It is this expression
for the neutrino energies that should be used in the averaged energy-momentum
conservation equation (\ref{neutrino}) that characterises a scattering
event between a neutrino and a D-particle. From (\ref{LI}), then, we obtain that 
the total combined effect on the energy-momentum dispersion relations,
from both capture/splitting and metric distortion, can be represented
as:
$\ll E_{\rm after}\gg=\pm\sqrt{p^{2}+m^{2}}\left(1+\frac{1}{2}\sigma^{2}\right)-\frac{1}{2}\frac{M_{s}}{g_{s}}\,\sigma^{2}~.$
Since antiparticles of spin 1/2 fermions can be viewed as ``holes''
with negative energies, we obtain from (\ref{neutrino}) and (\ref{dispersion3})
the following dispersion relations between particles and antiparticles
in this geometry (for Majorana neutrinos, the r\^oles of particles /antiparticles
are replaced by left/right handed fermions):
\begin{eqnarray}
\ll E_{\nu}\gg  =  \sqrt{p^{2}+m_{\nu}^{2}}\left(1+\frac{1}{2}\sigma^{2}\right)-\frac{1}{2}\frac{M_{s}}{g_{s}}\,\sigma^{2}~, \quad 
\ll E_{\overline{\nu}}\gg  =  \sqrt{p^{2}+m_{\nu}^{2}}\left(1+\frac{1}{2}\sigma^{2}\right)+\frac{1}{2}\frac{M_{s}}{g_{s}}\,\sigma^{2}\label{cptvdisp}
\end{eqnarray}
where $E_{\overline{\nu}}>0$ represents the positive energy of a physical
antiparticle. In our analysis above we have made the symmetric assumption
that the recoil-velocities fluctuation strengths are the same between
particle and antiparticle sectors (scenarios for which this symmetry
was not assumed have also been considered in an early work~\cite{Bernabeu:2006av}).
There can thus be \emph{local} CPTV in the sense that the effective
dispersion relation between neutrinos and antineutrinos are different.
This is a consequence of the local violation of Lorentz symmetry (LV),
as a result of the non-trivial recoil velocities of the D-particle,
leading to the LV space-time distortions (\ref{recmetric}).

The difference (\ref{cptvdisp}) in the dispersion relations 
between particles and antiparticles  will imply differences in the
relevant populations of neutrinos ($n$) and antineutrinos (${\overline{n}}$). This difference between neutrino
and antineutrino phase-space distribution functions in D-foam backgrounds
generates a matter-antimatter lepton asymmetry in the relevant densities
\begin{equation}
\ll n-{\overline{n}}\gg=g_{d.o.f.}\int\frac{d^{3}p}{(2\pi)^{3}}\ll[f(E)-f(\overline{E})]\gg~,\quad f(E,\mu)=\frac{1}{{\rm exp}(E-\mu)/T)\pm1}~,\quad E^{2}=\vec{p}^{2}+m^{2}\
\label{Leptonasym}
\end{equation}
 where $g_{d.o.f.}$ denotes the number of degrees of freedom of relativistic
neutrinos, and $\ll\dots\gg$ denotes an average over suitable populations
of stochastically fluctuating D-particles (\ref{LI}).

For the purposes of this talk, we shall make the plausible simplifying assumption that $\sigma^{2}$ is constant
\emph{i.e.} independent of space and of the
(anti)neutrino energy. It is a parameter which can only be \emph{positive}.
This is for estimation purposes only. A more detailed and complete analysis will be given elsewhere~\cite{sarbennick}. 
Ignoring neutrino
mass terms and $(1+\frac{\sigma^{2}}{2})$ square-root prefactors
in (\ref{cptvdisp}), setting the (anti)neutrino chemical potential
to zero (which is a sufficient approximation for relativistic light
neutrino matter) and performing a change of variables $|\vec{p}|/T\to\tilde{u}$
we obtain from (\ref{Leptonasym}) the result:
\begin{eqnarray}\label{dn2}
\Delta n_{\nu} & = & \frac{g_{d.o.f.}}{2\pi^{2}}\, T^{3}\int_{0}^{\infty}d\tilde{u}\,\tilde{u}^{2}\,[\frac{1}{1+e^{\tilde{u}-\frac{M_{s}\sigma^{2}}{2g_{s}\, T}}}-\frac{1}{1+e^{\tilde{u}+\frac{M_{s}\sigma^{2}}{2\, g_{s}\, T}}}]=\frac{g_{d.o.f.}}{\pi^{2}}\, T^{3}\left({\rm Li}_{3}(-e^{-\frac{M_{s}\sigma^{2}}{2g_{s}\, T}})-{\rm Li}_{3}(-e^{\frac{M_{s}\sigma^{2}}{2g_{s}\, T}})\right)\nonumber \\
 & \simeq & \frac{g_{d.o.f.}}{\pi^{2}}\, T^{3}\left(\frac{M_{s}\sigma^{2}}{g_{s}\, T}\right)\,>\,0,
\end{eqnarray}
 to leading order in $\sigma^{2}$, where in the last step we took
into account the formal definition as a series of the Polylogarithm
function ${\rm Li}_{s}(z)=\sum_{k=1}^{\infty}\,\frac{z^{k}}{k^{s}}$
which is valid for $|z|\,<\,1$, while the cases $|z|\ge1$ are defined
by analytic continuation. We thus observe that the CPTV term $-\frac{1}{2}\frac{M_{s}}{g_{s}}\sigma^{2}$
in the dispersion relation (\ref{cptvdisp}) for the neutrino, which
corresponds to the energy `loss' due to the D-particle recoil kinetic
energies, comes with the right sign (\emph{`loss}') so as to guarantee an
\emph{excess} of particles over antiparticles. This is a nice feature of our string model, which is not met in other CPTV cases, induced by the coupling of the fermion (neutrino) spin to local curvature in  (axisymmetric) background space-times, that could characterise the early Universe~\cite{mukho}, where the positive sign of $\Delta n$ has to be fixed by hand. 

As in standard scenarios of Leptogenesis, the Lepton asymmetry (\ref{dn2}), $\Delta n_\nu/s$, with $s \propto T^3$, 
decreases with decreasing temperature up to a freeze-out point, which
occurs at temperatures $T_{d}$ at which the Lepton-number violating
processes decouple. This is taken to be the conventional one (in standard
scenarios of Leptogenesis): $T_{d}\sim10^{15}$~GeV. The resulting lepton asymmetry then freezes out to a value:
\begin{equation}
\Delta L(T<T_{d})=\frac{\Delta n_{\nu}}{s}\sim\frac{M_{s}\,\sigma^{2}}{g_{s}\, T_{d}}\label{Ddl}
\end{equation}
 which survives today. The so calculated $\Delta L$ assumes the phenomenologically
relevant order of magnitude of $10^{-10}$ provided
$\frac{M_{s}}{g_{s}}\,\sigma^{2}\sim10^{5}\,{\rm GeV}.$
Since in these scenarios, the dimensionless stochastic variable,
expressing fluctuations of a recoil velocity, is always less than
one, $\sigma^{2}<1$, one observes that the required Lepton asymmetry
is obtained for D-particle masses larger than
$\frac{M_{s}}{g_{s}}\,>\,100~{\rm TeV}.$ 
The so obtained $\Delta L$ can then be communicated to the baryon
sector, to yield the observed (today) baryon asymmetry (\ref{basym}), via either
B+L violating sphaleron processes~\cite{cpbau}, or B-L conserving interactions
in Grand Unified theories, as in standard scenarios.

\section{Conclusions and Outlook}\label{sec:3}
In this talk, taking advantage of the theme of the conference on New Frontiers in Physics, I have discussed some rather unconventional 
features of string-inspired cosmologies associated with the non-trivial r\^ole of moduli fields in inducing 
modifications in the Cosmology sector of string theories with profound implications in their low-energy phenomenology. 
Specifically, I have discussed how \emph{running} (cosmic-time dependent) \emph{dilaton} scalar fields in the gravitational multiplet of strings, can affect DM thermal relic abundances by diluting them, compared to the standard (dilaton-free) cosmology case. This leads to an enlargement of the  cosmology-allowed parameter space of minimal supersymmetric models that can be accommodated in the low-energy limit of such string theories. Depending on the magnitude of the dilution, the SUSY partner masses in mSUGRA-type models can be pushed to higher values, and in some cases even beyond the reach of the full 14 TeV run of the LHC.
Thus, a simple extension of the string-inspired cosmology, by allowing a time-dependent dilaton field in the Early Universe (before BBN), suffices to account for the absence of any signature of minimal SUSY in the current  LHC run and may be in future colliders. Of course the cosmology of such running dilatons is complicated and highly model dependent, and comparison with current astrophysical and cosmological data is expected to rule out several of these models. Nevertheless, at present one can find models consistent with the data, including those on structure formation in the Universe, if one assumes that running dilatons are present at such late epochs (we stress, though, that this is not necessary, and models exist where the running-dilaton dominance ends before BBN). 

As a second interesting topic, we have discussed how in the above models CPT may be violated in the early Universe in the presence of cosmic, effectively `point-like'  brane defects (``D-particles''), which could be abundant at early eras but not today. The model invokes brane worlds and the space-time defects can populate both the brane and the higher-dimensional bulk space, in which the brane worlds are embedded. The interaction of matter open strings with such defects on the brane worlds causes, due to defect recoil, \emph{local} violations of Lorentz symmetry, which may, thus, be conserved only on average over statistically significant populations of defects. This causes CPT Violation in the sense of different dispersion relations induced between particles and antiparticles, when moving in the ``medium'' of these defects. There is a preferential r\^ole of neutrinos, over other elementary excitations of the SM, in feeling the full effects of the D-particle medium, which has been attributed to electric charge conservation, since the defects are electrically neutral. The resulting differences between neutrino/antineutrino in the effective dispersion relations lead, under natural assumptions, to a Lepton asymmetry at early epochs of the Universe, at neutrino decoupling temperatures of $10^{15}$~GeV, which has the correct phenomenological value to lead to the observed Baryon Asymmetry today via Baryon-Lepton number violating processes in grand unified theories of the low-energy limit of string models. The advantage of this type of defect-induced CPT Violation is that it does not require details of supersymmetry breaking nor extra sources of CP violation, such as sterile neutrinos.

We close by stating once more that it remains to be seen whether String Cosmology offers the true description of the physical Universe. As we have mentioned in the introduction, there are still many open problems in string theory, the Landscape being in my opinion one of the most important of them (although others do not seem to consider it as an issue and invoke anthropic arguments in order to `select' the physical (observed) ground state from the plethora of string vacua). Nevertheless, progress is being made and hopefully one day the fundamental (non-perturbative) structure of string/brane theory at microscopic scales will be understood. It is only then that concrete falsifiable predictions can be made. Until that moment, in order to make contact with the real world,  we should restrict ourselves to low-energy phenomenological studies such as the ones presented in this talk, which however, as  I hope I convinced you, are of great value and worthy of pursuing.

\begin{acknowledgement}

I thank the organisers of the 1st ICFP 2012 in Kolymbari (Crete, Greece) for their invitation.  This work is  supported in  part by  the London  Centre for
Terauniverse Studies (LCTS), using  funding from the European Research
Council  via  the  Advanced  Investigator  Grant 267352,  and  by STFC (UK) 
research grant ST/J002798/1. 

\end{acknowledgement}

\end{document}